\documentstyle[preprint,aps]{revtex}

\tighten
\input epsf

\begin{document}
\draft

\title{\Large \bf Microscopic Deterministic Dynamics and
 Persistence Exponent}

\author{\bf B. Zheng}

\address{FB Physik, Universit\"at -- Halle, 06099 Halle, Germany}

\maketitle

\begin{abstract}
Numerically we solve the microscopic deterministic equations
of motion with random initial states
for the two-dimensional $\phi^4$ theory. 
Scaling behavior of the persistence probability
at criticality is
systematically investigated and
the persistence exponent is estimated.
\end{abstract}


Recently the persistence exponent has attracted much attention.
This exponent was first introduced in the context of
the non-equilibrium coarsening dynamics at zero temperature
\cite {der94,bra94b}. It characterizes the power law decay
of the {\it persistence} probability that a {\it local} order parameter
keeps its sign during a time $t$ after
a quench from a very high temperature to zero temperature.
 For {\it critical} dynamics, the local order parameter
(usually, a spin) flips rapidly and the persistence probability
does not obey a power law. 
In this case, however, the persistence exponent
$\theta_p$ can be defined
by the power law decay of the {\it global} persistence probability
$p(t)$
 that the global order parameter has not changed the sign in a
 time $t$ after the quench from a very high temperature to the
 critical temperature \cite {maj96a},
\begin{equation}
p(t) \sim \, t^{\theta_p}\ .
\label{e10}
\end{equation}

An interesting property of the persistence exponent
is that its value is highly non-trivial even for simple systems.
For the quench to zero temperature, 
for example, $\theta_p$ is apparently not a simple fraction
for the simple diffusion equation and
the Potts model in one dimension
\cite {der95,maj96}.
For the quench to the critical temperature,
it is shown that the persistence exponent
is generally a new {\it independent} critical exponent,
i.e. it can not be expressed by the known static exponents,
the dynamic exponent $z$ and the recently discovered
 exponent $\theta$ \cite {maj96a,oer97}. This relies on the fact that
 the time evolution
of the global magnetization is {\it not} a Markovian process.
Recent Monte Carlo simulations for the 
Ising and Potts model at criticality 
support the power law decay of the global 
persistence probability and detect also
the non-Markovian effect \cite {sta96,sch97}.

Up to now, the persistence exponent has been studied
only in {\it stochastic} dynamic systems, described typically by
Langevin equations or Monte Carlo algorithms.
From fundamental view points, both equilibrium
 and non-equilibrium properties
of statistical systems can be described by the 
{\it microscopic
deterministic} equations of motion
(e.g. Newton, Hamiltonian and Heisenberg equations)
\footnote {Langevin equations at zero temperature
are also deterministic, but they are at mesoscopic level
and generally different from the microscopic
deterministic equations of motion.}, even though a general proof
does not exist. With recent development of computers,
 gradually it becomes possible to solve numerically 
the microscopic
deterministic equations of motion.
For example, the $O(N)$ vector model
 and the $XY$ model have been investigated
 \cite {cai98,cai98a,leo98}.
 The results confirm that the deterministic equations 
 describe correctly second order phase transitions.
 The static critical exponents are estimated
 and agree with existing values.
 More interestingly, in a recent paper
 short-time dynamic behavior of the deterministic dynamics
starting from
 random initial states has been studied  \cite {zhe99}. 
 The short-time dynamic scaling was found
 and the estimated value of the dynamic exponent $z$ 
 is the same as that of
  the Monte Carlo dynamics
 of the Ising model.
 
 The purpose of this letter is to study 
 the {\it critical} scaling behavior of the global
 persistence probability and measure the persistence exponent
 in microscopic deterministic dynamic systems,
 taking the two-dimensional $\phi^4$ theory
 as an example.
 
 The Hamiltonian of the two-dimensional $\phi^4$ theory 
 on a square lattice is written as
\begin{equation}
H=\sum_i \left [ \frac{1}{2} \pi_i^2 
  + \frac{1}{2} \sum_\mu (\phi_{i+\mu}-\phi_i)^2 
  - \frac{1}{2} m^2 \phi_i^2 
  + \frac{1}{4!} \lambda \phi_i^4 \right ]
\label{e20}
\end{equation}
with $\pi_i=\dot \phi_i$ and it leads to the equations of motion
\begin{equation}
 \ddot \phi_i= \sum_\mu (\phi_{i+\mu}+\phi_{i-\mu}- 2\phi_i)
  +  m^2 \phi_i
  - \frac{1}{3!} \lambda \phi_i^3\ .
\label{e30}
\end{equation}

 Energy is conserved
during the dynamic evolution governed by
 Eq. (\ref {e30}). 
 As discussed in Refs. \cite {cai98,zhe99},
  a microcanonical ensemble
 is assumed to be generated by the solutions.
 In this case, the temperature could not be introduced 
externally as in a canonical ensemble,
but could only be defined internally as the averaged
kinetic energy.
In the dynamic approach,
the total energy is actually 
an even more convenient controlling parameter
of the system, since it is conserved and 
can be input from the initial state. 

From the view point of ergodicity, to achieve
a correct equilibrium state the microscopic deterministic
dynamic system should
start from a {\it random} initial state.
Interestingly, this is just similar  to
the dynamic relaxation in stochastic dynamics
after a quench from a very high temperature.
Therefore, similar dynamic behavior may be expected
for both dynamic systems. 

The order parameter of the $\phi^4$ theory is the magnetization
$M(t)=\sum_i \phi_i(t)/L^2$
with $L$ being the lattice size.
In this paper, we are interested in
 the global persistence probability $p(t)$ at the critical point,
 which is defined as the probability
 that the {\it not averaged} order parameter 
 has not changed the sign in a time $t$ 
 starting from a random state 
 with small initial magnetization $m_0$.
 
Following Ref. \cite {cai98,zhe99}, we take parameters $m^2=2.$
and $\lambda=0.6$ and prepare the initial configurations as follows.
For simplicity, we set initial kinetic energy to be
zero, i.e. $\dot \phi_i(0)=0$. We fix
the magnitude of the initial field to be a constant $c$, 
$|\phi_i(0)|=c$, and then randomly give the sign to $\phi_i(0)$
with the restriction of a fixed magnetization in unit of $c$,
and finally the constant $c$ is determined by
the given energy.

To solve the equations of motion (\ref {e30}) numerically,
 we simply discretize
$\ddot \phi_i$ by 
$(\phi_i(t+\Delta t)+\phi_i(t-\Delta t)-2\phi_i(t))/(\Delta t)^2$.
According to the experience 
in Ref. \cite {zhe99}, $\Delta t$ is taken to be $0.05$.
After an initial configuration
is prepared,
we update the equations of motion until
the magnetization changes its sign.
The maximum observing time is
$t=1000$. Then we repeat the procedure with other
initial configurations and measure the persistence probability $p(t)$.
In our calculations, we use 
fairly large lattices $L=128$ and $256$ and samples
of initial configurations
 range from $3\ 500$ to $30\ 000$
depending on initial magnetization $m_0$ and lattice sizes.
The smaller $m_0$ and lattice size $L$ are,
the more samples of initial configurations we have.
Errors are simply estimated by dividing total samples
into two or three subsamples.
Compared with Monte Carlo simulations,
our calculations here are much more time consuming
due to the small $\Delta t$.

 According to analytical analyses and Monte Carlo simulations
 in stochastic dynamic systems, 
 at the critical point and in the limit $m_0=0$,
$p(t)$ should decay by a power law as in Eq. (\ref {e10}).
Our first effort is to investigate
whether in microscopic deterministic dynamics $p(t)$ obeys also the power law 
and measure the persistence exponent $\theta_p$.

\begin{figure}[p]\centering
\epsfysize=6.cm
\epsfclipoff
\fboxsep=0pt
\setlength{\unitlength}{0.6cm}
\begin{picture}(9,9)(0,0)
\put(-2,-0.5){{\epsffile{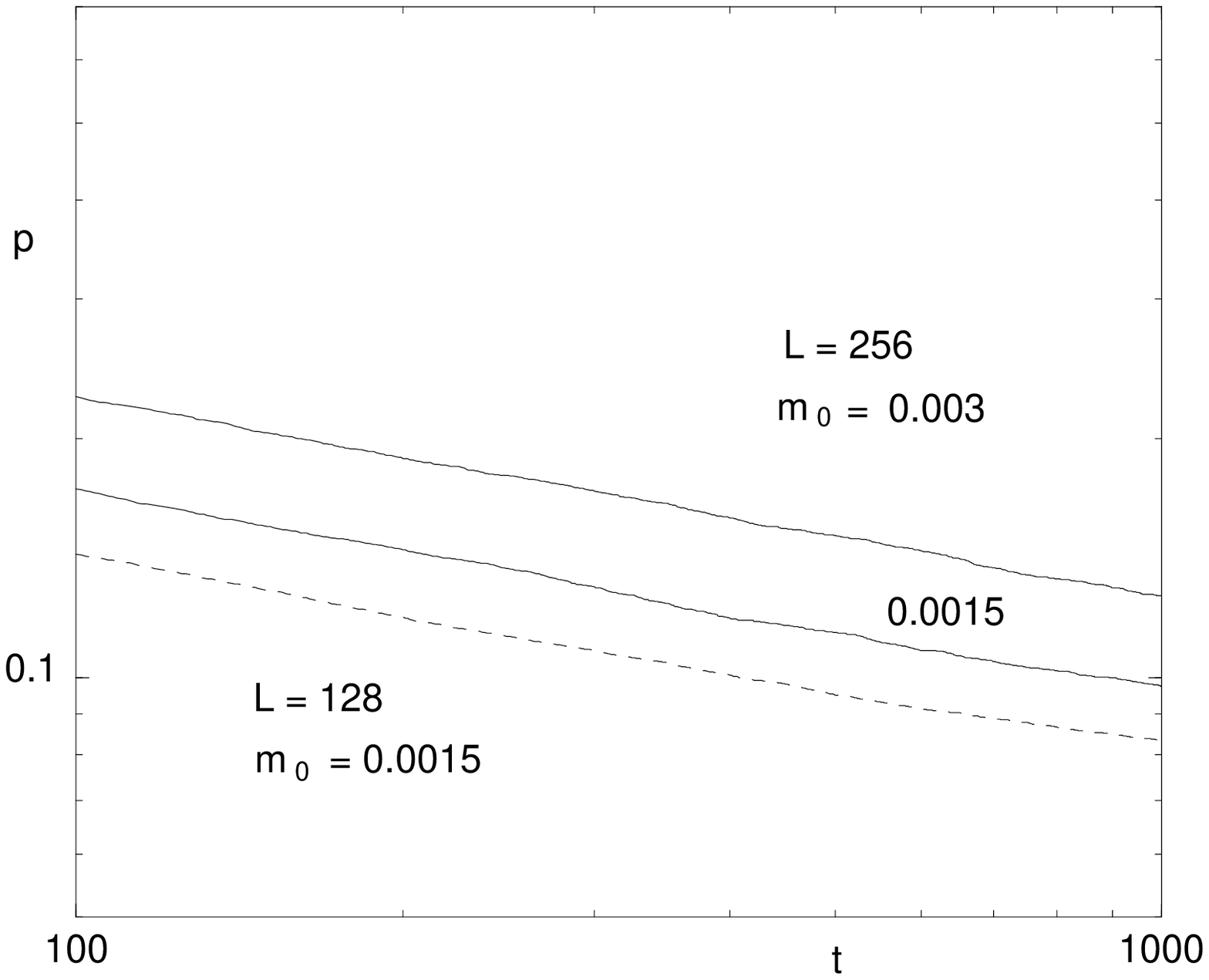}}}
\end{picture}
\caption{The persistence probability is displayed
in log-log scale for $\epsilon_c = 21.1$.
Solid lines and a dashed line are for lattice sizes $L=256$ and $128$
respectively.
}
\label{f1}
\end{figure}

Here we adopt the critical energy density
$\epsilon_c = 21.1$ from the literature \cite {cai98,zhe99}.
In Fig. \ref {f1}, the persistence probability $p(t)$
is displayed on a log-log scale for lattice sizes
$L=256$ and $128$ with solid lines and a dash line respectively.
For $L=256$ simulations have been performed with two 
values of initial magnetization
$m_0=0.003$ and $0.0015$.
These straight lines convince us the power law 
behavior of $p(t)$.
 Skipping data within a microscopic time scale 
 $t \sim 100$, from the slopes of the curves one estimates the
 persistence exponent $\theta_p=0.252(6)$
 for both values of $m_0$.
 This shows that there is already no effect of finite $m_0$. 
 The curve for $L=128$ and $m_0=0.0015$ is roughly parallel to
 that of $L=256$. One measures the the slope $\theta_p=0.251(1)$ 
 in the time interval $[100,500]$
 but $\theta_p=0.232(10)$ in $[100,1000]$.
 This indicates that some finite size effect exists still
 for $L=128$ after $t=500$ but is negligible small for $L=256$.
 
If the time evolution of the magnetization is a Markovian process,
from theoretical view points 
the persistence exponent will be not an independent critical exponent
and it will take the value $\alpha_p$, which relates
to other exponents through
\begin{equation}
\alpha_p = - \theta + (d/2-\beta/\nu)/z \ .
\label{e40}
\end{equation}
In Table \ref {t1}, values of the exponent $\theta_p$,
$z$, $\theta$ and $\alpha_p$ for the $\phi^4$ theory
are given in comparison with those of the kinetic Ising model
induced by local Monte Carlo algorithms.
As is the case of the Ising model, the exponents 
$\theta_p$ and $\alpha_p$ for the $\phi^4$ theory
differ also by about $10$ percent.
This represents a rather visible non-Markovian effect
in the time evolution of the magnetization.

For equilibrium states, it is generally believed that
the $\phi^4$ theory and the Ising model are 
in a same universality class.
Results from numerical solutions of the deterministic equations
also support this \cite {cai98}.
From the short-time dynamic approach \cite {zhe99},
within statistical errors
the dynamic exponent $z$ for the microscopic
deterministic dynamics
of the $\phi^4$ theory is
the same as that of the kinetic Ising model
with Monte Carlo algorithms
but the exponent $\theta$ differs by several percent.
In Table \ref {t1}, we see that  
$\theta_p$ for the $\phi^4$
theory is also several percent 
bigger than that of the Ising model.
However, by feeling we still think that the $\phi^4$
theory and the Ising model are very probably
in a same persistence
universality class. These some percent differences
of the exponents come probably from that
the critical point $\epsilon_c$ has not been very accurate
or there are some corrections to scaling
\footnote {In deterministic dynamics, 
energy is conserved and it couples to the order parameter.
Therefore, it is believed that 
the deterministic dynamics belongs 
to the dynamics of model C rather than model A.
Standard local  Monte Carlo dynamics of the Ising model is 
dynamics of model A. In two dimensions, model A and C
are the same but maybe up to a logarithmic correction
\cite {hoh77,oer99}.}
and uncontrolled systematic errors.
Actually, we will see below that 
the critical energy density could be somewhat
lower than $\epsilon_c=21.1$ 
and it would yield a slightly smaller $\theta_p$.

\begin{figure}[p]\centering
\epsfysize=6.cm
\epsfclipoff
\fboxsep=0pt
\setlength{\unitlength}{0.6cm}
\begin{picture}(9,9)(0,0)
\put(-2,-0.5){{\epsffile{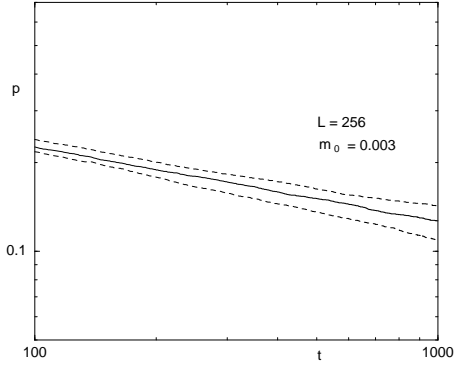}}}
\end{picture}
\caption{The persistence probability is displayed
in log-log scale
for the energy densities $\epsilon=20.9$, $21.1$ and
$21.3$ (from above).
}
\label{f2}
\end{figure}

Our second step is to investigate the scaling behavior
of the persistence probability  in the neighborhood
 of the critical energy density.
 From general view points of physics,
 one may expect a following scaling form 
 \begin{equation}
p(t,\tau)= t^{-\theta_p} F(t^{1/\nu z}\tau) \ .
\label{e50}
\end{equation}
Here $\tau=(\epsilon-\epsilon_c)/\epsilon_c$
is the reduced energy density.
When $\tau=0$, the power law behavior in Eq. (\ref {e10})
is recovered. When $\tau$ differs from zero,
the power law will be modified by the scaling function
$F(t^{1/\nu z}\tau)$.
In principle, this fact may be used for the determination
of the critical energy density.
In Fig.~\ref {f2}, $p(t,\tau)$ for $L=256$ and $m_0=0.003$
is plotted for three different
energy density $\epsilon=20.9$, $21.1$ and
$21.3$. 
In the figure, we see that the solid line shows the 
best power law behavior among the three curves. 
However, an very accurate
estimate of $\epsilon_c$ could not be achieved so easily,
since $p(t,\tau)$ is not so sensitive to the energy.
According to our data, we estimate
$\epsilon_c=21.06(12)$. Within errors, it is consistent
with $\epsilon_c=21.1$ given in \cite {cai98}
and $\epsilon_c=21.11(3)$ in \cite {zhe99}.
We should point out, that the exponent $\theta_p$ will be $0.245$,
if it is measured at $\epsilon_c=21.06$.
It is closer to that of the kinetic Ising model,
as discussed above.

\begin{figure}[p]\centering
\epsfysize=6.cm
\epsfclipoff
\fboxsep=0pt
\setlength{\unitlength}{0.6cm}
\begin{picture}(9,9)(0,0)
\put(6.,4.){\makebox(0,0){\footnotesize $\partial_\tau \ \ln \ p$}}
\put(-2,-0.5){{\epsffile{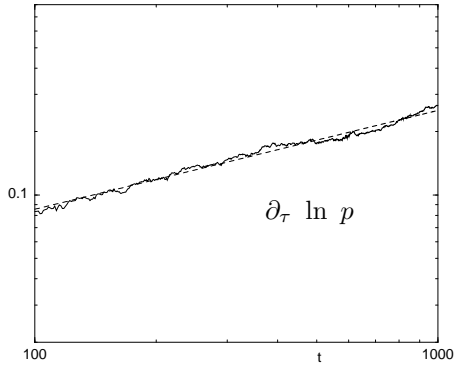}}}
\end{picture}
\caption{The logarithmic derivative of the persistence probability
with respect to the energy density is plotted
in log-log scale, using the data in Fig.~\protect\ref {f2}.
The dashed line is the best fitted straight line
}
\label{f3}
\end{figure}

In order to have more understanding of
the scaling form (\ref {e50}),
we differentiate with respect to the energy density
on both sides of the equation and obtain
\begin{equation}
\partial_{\tau} \ \ln p(t,\tau) \ |_{\tau=0} \sim t^{1/\nu z}  \ .
\label{e60}
\end{equation}
Using the data of Fig.~\ref {f2}, we can approximately
calculate $\partial_{\tau} \ \ln p(t,\tau) |_{\tau=0}$
and the result is displayed in Fig.~\ref {f3}.
Even though there are some fluctuations, power law behavior
is still seen. The best fitted slope of the curve gives
$1/\nu z = 0.47(4)$ in the time interval
$[100,1000]$. Taking $z=2.15(2)$ as input,
one obtains $\nu=0.99(8)$.
Compared with $\nu=1$ for the Ising model,
this result supports that the $\phi^4$ theory
with deterministic dynamics and 
the Ising model are in a same universality class.  

Finally we study the scaling behavior of the persistence probability 
in case the initial magnetization is not so small and its effect
can not be neglected.  
Following Ref.~\cite {sch97}, we assume a finite size scaling form
\begin{equation}
p(t,L,m_0)= t^{-\theta_p} F(t^{1/z}L^{-1},t^{x_{0p}/z}m_0) \ .
\label{e70}
\end{equation}
Here the energy density has been set to its critical value
and $x_{0p}$ is the scaling dimension of
the initial magnetization $m_0$.
It was discovered that for the Ising model with Monte Carlo dynamics,
the value $x_{0p}=1.01(1)$ is 'anomalous',
i.e. it is different from the scaling dimension 
of the initial magnetization $x_0=0.536(2)$ 
measured from the time evolution of
the magnetization or auto-correlation \cite {sch97}.
The origin should be 
that $p(t,L,m_0)$ is a {\it non-local} observable in
 time $t$. It remembers the history of the time evolution.

\begin{figure}[p]\centering
\epsfysize=6.cm
\epsfclipoff
\fboxsep=0pt
\setlength{\unitlength}{0.6cm}
\begin{picture}(9,9)(0,0)
\put(-2,-0.5){{\epsffile{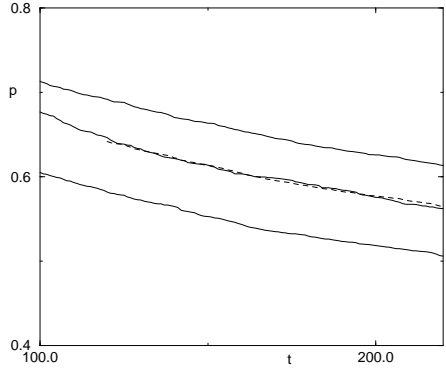}}}
\end{picture}
\caption{The scaling plot for the persistence probability.
The lower and upper solid lines are the persistence probability
for $L_2=128$ with $m_{0}=0.0272$ and $0.0349$ respectively,
while the dashed line is the properly rescaled one
for $L_1=256$ with $m_{01}=0.0151$.
The solid line fitted to the dashed line
represents the persistence probability
for $L_2=128$ with $m_{02}=0.0313(3)$.
}
\label{f4}
\end{figure}

To verify the scaling form (\ref {e70}) and
estimate $x_{0p}$, we perform a simulation with 
the lattice size $L_1=256$ and initial magnetization
$m_{01}=0.0151$. Suppose the scaling form (\ref {e60}) holds,
one can find an initial magnetization $m_{02}$ with the lattice size
$L_2=128$ such that the curves of $p(t,L,m_0)$
for both lattice sizes collapse.
Practically we haved performed 
simulations for $L_2=128$ with two initial magnetizations, 
$m_0=0.0272$ and $0.0349$.
By linear extrapolation, we obtain data with $m_0$ between
$0.0272$ and $0.0349$.
Searching for a curve best fitted to the curve
for $L_1=256$, we determine $m_{02}$.
In Fig.~\ref {f4}, such a scaling plot is displayed.
The lower and upper solid lines are the persistence probability
for $L_2=128$ with $m_0=0.0272$ and $0.0349$ respectively,
while the dashed line is the properly rescaled one
for $L_1=256$ with $m_{01}=0.0151$.
The solid line fitted to the dashed line
represents the persistence probability
for $L_2=128$ with $m_{02}=0.0313(3)$.
Since the microscopic time scale is $t_{mic} \sim 100$,
nice collapse of the two curves can be observed only
after $t \sim 100$.
From the scaling form (\ref {e70}),
$m_{02}=2^{x_{0p}}\ m_{01}$ and one estimates
$x_{0p}=1.05(1)$. This value is very close to
 $x_{0p}=1.01(1)$ for the kinetic Ising model.

 In conclusions, we have numerically solved 
 the microscopic deterministic equations of
 motion with random initial states
 for the two-dimensional $\phi^4$ theory
 and systematically investigated 
 the critical scaling behavior of the persistence probability.
As summarized in Table~\ref {t1}, the estimated 
exponents $\theta_p$ and $x_{0p}$
are very close to those of the kinetic Ising model
induced by local Monte Carlo algorithms.
 What would be the dynamic and static 
 behavior of the deterministic dynamics
 starting from more general initial states is
 an interesting work in future.

{\bf Acknowledgements}:
Work supported in part by the Deutsche Forschungsgemeinschaft
under the project TR 300/3-1.

\begin{table}[h]\centering
\begin{tabular}{cccccccc}
         & $\theta_p$  & $z$       &   $\theta$     & $\alpha_p$
                 & $1/\nu z$          & $\nu$       & $x_{0p}$     \\
$\phi^4$ & 0.252(6)    & 2.15(2)   &    0.176(7)    &  0.231(7)
                 & 0.47(4)         &    0.99(8)     &  1.05(1) \\
Ising  & 0.238(3)& 2.155(3)        &    0.191(1)    &  0.215(1)
                 &                 &      1         &  1.01(1)  \\
\end{tabular}
\caption{The critical exponents measured for the $\phi^4$ theory
in comparison with those of the Ising model.
The value of $\nu$ for the Ising model
is exact, while others are taken from Table 2 in
Ref.~\protect\cite {zhe98} and from Ref.~\protect\cite {sch97}.
$z$ and $\theta$ for the $\phi^4$ theory are 
from Ref.~\protect\cite {zhe99}.
To calculate $\alpha_{p}$ for the $\phi^4$ theory $\beta/\nu=1/8$
is taken as input.
}
\label{t1}
\end{table}


\begin{thebibliography}{10}

\bibitem{der94}
{B. Derrida, A.J. Bray and C. Godr\'eche}, J. Phys. {\bf {A27}},  L357  (1994).

\bibitem{bra94b}
{A. J. Bray, B. Derrida and C. Godr\'eche}, Europhys. Lett. {\bf {27}},  175
  (1994).

\bibitem{maj96a}
{S.N. Majumdar, A.J. Bray, S. Cornell and C. Sire}, Phys. Rev. Lett. {\bf
  {77}},  3704  (1996).

\bibitem{der95}
{B. Derrida, V. Hakim and V. Pasquier}, Phys. Rev. Lett. {\bf {75}},  751
  (1995).

\bibitem{maj96}
{S.N. Majumdar, C. Sire, A.J. Bray and S. Cornell}, Phys. Rev. Lett. {\bf
  {77}},  2867  (1996).

\bibitem{oer97}
{K. Oerding, S.J. Cornell and A.J. Bray}, Phys. Rev. {\bf {E56}},  R25  (1997).

\bibitem{sta96}
{D. Stauffer}, Int. J. Mod. Phys. {\bf {C7}},  753  (1996).

\bibitem{sch97}
L. {Sch\"ulke} and B. Zheng, Phys. Lett. {\bf A 233},  93  (1997).

\bibitem{cai98}
{L. Caiani, L. Casetti and M. Pettini}, J. Phys. {\bf {A31}},  3357  (1998).

\bibitem{cai98a}
{L. Caiani, L. Casetti, C. Clementi, G. Pettini, M. Pettini and R. Gatto},
  Phys. Rev. {\bf {E57}},  3886  (1998).

\bibitem{leo98}
{X. Leoncini and A.D. Verga}, Phys. Rev. {\bf {E57}},  6377  (1998).

\bibitem{zhe99}
{B. Zheng, M. Schulz and S. Trimper}, Phy. Rev. Lett. {\bf {82}},  1891
  (1999).

\bibitem{hoh77}
{P.C. Hohenberg and B.I. Halperin}, Rev. Mod. Phys. {\bf {49}},  435  (1977).

\bibitem{oer99}
{K. Oerding}, {\em {private communication}}, 1999.

\bibitem{zhe98}
B. Zheng, Int. J. Mod. Phys. {\bf B12},  1419  (1998), review article.

\end{thebibliography}
\end{document}